\documentclass[twocolumn,showpacs,preprintnumbers,amsmath,amssymb]{revtex4-1}
\usepackage{amsmath}
\usepackage{amssymb}

\usepackage{graphicx}
\usepackage{dcolumn}
\usepackage{bm}
\usepackage{color}

\begin{document}

\title{Thermal transport due to quantum interference in magnetic tunnel junctions }
\author{J. F. Feng, D. P. Liu, Q. L. Ma, H. X. Wei, and X. F. Han}
\affiliation{State Key Laboratory of Magnetism, Beijing National Laboratory for Condensed Matter Physics, Institute of Physics, Chinese Academy of Sciences, Beijing 100190, China}
\date{\today}

\begin{abstract}
We study the thermal transport in magnetic tunnel junctions. Thermal gradients across the tunneling barrier appear around the Fowler-Nordheim tunneling regime, due to the current-induced heat caused by quantum interference. Both thermovoltage and thermal temperature follow a linear response with the applied current, which is an evidence for a thermoelectric effect. By increasing the barrier transparency, the dynamics of thermoelectric properties is observed with the current. Accordingly, a large range of the Seebeck coefficient, 10 - 1000 $\mu V/K$, has been obtained in magnetic tunnel junctions.

\end{abstract}



\maketitle

\section{Introduction}

Thermoelectric phenomena have been discovered by T.J. Seebeck one
century ago. Recently, people studied thermoelectronics in
spintronic devices to explore the basic physics and potential
applications \cite{1JShi, 2MHatami, 3KUchida, 4GEWBauer, 5MWalter,
6ASlachter, 7MCzerner, 8NLiebing, 9JFlipse, 10WWLin, 11HMYu,
12ASlachter, 13GEWBauer}. The coupling of thermoelectronics with
spintronics has generated novel research fields, such as
thermoelectric effect \cite{7MCzerner, 8NLiebing, 9JFlipse, 10WWLin},
thermal spin transfer torque \cite{11HMYu}, and thermally driven
spin injection \cite{12ASlachter}, etc. Magnetic tunnel junctions
(MTJs), spin valve structures, and ferromagnet/metal contact usually
serve as probes to detect thermoelectronics. Furthermore, a heat
source, e.g. a dielectric material of $AlO_{x}$, is needed to
generate a thermal gradient across the probe device \cite{7MCzerner,
8NLiebing, 9JFlipse, 10WWLin}. Besides, the nonlocal method is also
used to detect the thermoelectronic phenonmena \cite{12ASlachter}. Until now,
only few reports have been used the dielectric material as
the heat source directly \cite{14ZHZhang}.

An MTJ is a sandwich structure with two ferromagnetic layers
separated by a thin insulating layer\cite{15MJulliere}. The
spin-dependent electronic transport in an MTJ is usually focused,
e.g., tunneling magnetoresistance (TMR) which is defined by
resistances in both parallel (PC) and anti-parallel configurations
(APC) of the ferromagnetic layers (TMR = $(R_{APC}-R_{PC})/R_{PC}$).
Previously, the voltage dependent TMR features have been explained
in terms of the spin-polarized band structure and \emph{ab-inito}
DFT calculations for a small voltage ($<$ $1V$)\cite{16XFHan,17DWang,18MSharma,19FMontaigne}. As increasing the
applied voltage, the barrier may become transparency for hot
electrons, which is out of the direct tunneling
regime\cite{20SZhang}. Moreover, the Fowler-Nordheim (FN) tunneling may occur when electrons enter the conduction band of the barrier \cite{21RFowler}.

As theoretically predicted\cite{22MBStearns,23AHDavies}, an oscillatory
TMR behavior occurs along with the FN tunneling, which is attributed
to the interferences of wave functions between ferromagnetic layers and the insulating barrier.
In this work, MTJs with an asymmetric Al-Oxide barrier have been fabricated to fit the
condition of the FN tunneling effect. We present
a systematic study of thermal transport of these MTJs
when the interference effect occurs\cite{24CWJBeenakker, 25OKarlstrom}. Here the tunneling barrier serves as the heat source directly and accordingly a thermal gradient across the barrier is
observed due to the current-induced heat.

Here we select the Seebeck effect to study such a phenomenon. The Seebeck effect is a fundamental phenomenon of thermoelectricity, which is a field subject to extensive research during the previous decades.\cite{26RDBarnard}. It normally deals with the interaction between heat transport and
the charge and spin degrees of freedom. In our case, the interference effect is mainly related to the charge of electron. Even though, from the spin polarization of MTJs, one could obtain the contributions of spin-up and spin-down electrons for Seebeck coefficient. It is proved that the method involved in this work
can explore thermoelectronics in MTJs \cite{27MWilczynski}.

\section{Experimental Details}

The MTJ samples in this work were grown by using an ultrahigh
vacuum (ULVAC) chamber of our magnetron sputtering system with a base pressure of
$1\times10^{-7}$ Pa\cite{28ZMZeng}. The sample structures have a bottom-to-top
sequence known as
IrMn(12)/$Co_{40}Fe_{40}B_{20}$(CoFeB)(4)/1.8 Al-oxide/CoFeB(3)
(thickness in nanometers). The Al-oxide barriers
were deposited by plasma oxidation with a mixture of oxygen and
argon at a pressure of 1.0 Pa in a separate chamber of the same sputtering system.
The $1.8nm$ barriers were fabricated in two steps to form an asymmetric
oxidation barrier: the first $0.9 nm$ Al layer was deposited with an
under-oxidation, and the second $0.9 nm$ Al was prepared
with a well optimized oxidation. All the samples were patterned into
ellipse-shaped junctions with the size of $(10\times12)\mu m^{2}$ by
conventional UV lithography. After this step, they were annealed at $260^{o}C$
under vacuum for one hour. All transport
measurements were performed with a four-probe technique at room
temperature. The positive voltage in this work is defined as
electron flow from bottom to top of MTJ stacks. Here we select two
MTJs labeled as MTJ1 and MTJ2 for a detailed study.

\section{Results and Discussion}


\begin{figure}
\includegraphics[width=8cm]{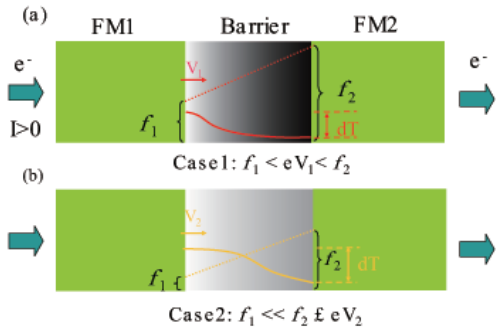}\\
\caption{Schematic of barrier heights ($\phi_{1}$ and $\phi_{2}$) and local thermal gradients
($\delta T_{1}$ at $V_{1}$ and $\delta T_{2}$ at $V_{2}$) across the tunneling barrier at the beginning of Case1 (a) and at the end of Case2 (b).}
\label{fig1}
\end{figure}

Because of the asymmetric feature of the tunneling barrier, the barrier height $\phi_{1}$ at the bottom
ferromagnet/barrier interface is lower than $\phi_{2}$ at the top
interface in these MTJs, as shown in Fig.1. Both $\phi_{1}$ and $\phi_{2}$ are higher than the
applied voltage at both PC and APC states in the low voltage range.
TMR varying with voltage mainly reflects density of states near the Fermi surface of the
ferromagnetic layers\cite{29MBowen}. With the increase of voltage, $\phi_{1}$ at two resistance states decreases, while $\phi_{2}$
increases, according to the fit results using the Brinkman¡¯s model\cite{30WFBrinkman}.
Once the applied voltage is close to or higher than $\phi_{1}$,
electrons appear in the barrier at the bottom interface side,
resulting in a FN tunneling. The oscillatory TMR with voltage is a
characteristic feature in the FN tunneling regime in MTJs \cite{31CWMiller, 32SHYang}; also see our data
shown in Fig. 2 (a). A negative TMR is clearly observed in a certain
positive voltage range (Fig. 2(c)). This suggests the FN tunneling is dominant (Case1).
With further increasing voltage, the transparency of the barrier
increases. Above a voltage, the barrier may behave like a spacer
partly, the spin-dependent scattering appears, which coexist with
the FN tunneling. After that, electrons may appear in the entire barrier with
more electrons at the bottom interface side. The spin-dependent
scattering enhances when the spacer - like barrier is formed (Case2 shown in Fig. 1 (b)).


\begin{figure}
\includegraphics[width=8cm]{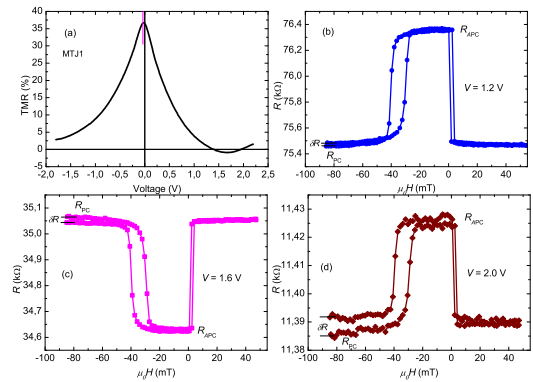}\\
\caption{(a) TMR as a function of voltage for MTJ1. (b)-(d) The junction resistance versus
magnetic field curves for MTJ1 at $1.2 V$, $1.6 V$, and $2.0 V$, respectively. The pink line in (a) indicates the offset from zero voltage, which may be responsible for the asymmetric feature of the barrier.}
\label{fig2}
\end{figure}

Now we discuss the heat effect in these MTJs. The interferences of the wave functions in the
conduction band of the barrier with electrodes at high voltage may
introduce some heat. From the current density dependence of exchange bias
($H_{ex}$) of MTJ1 as shown in Fig.3, $H_{ex}$
starts to decrease just before the FN tunneling occurs. The variation of exchange bias with current suggests the
spin-dependent tunneling current produces a heat, which can change
or revise the exchange bias \cite{33EGapihan}. At the mean while,
there is a local thermal gradient across the barrier \cite{33EGapihan}, which may produce the heat during the measurements. From the $R-\mu_{0}H$ curves shown in Fig.2 (b)- Fig.2 (d), a small resistance shift $(\delta R)$ is found.
This shift between the initial and final resistances suggests the heat appears in every
$R-\mu_{0}H$ loop, although the maximum current for MTJ1
is only $0.19 mA$ at $2.2 V$ (corresponding to a current density of $5.7
MA/m^{2}$).

\begin{figure}
\includegraphics[width=8cm]{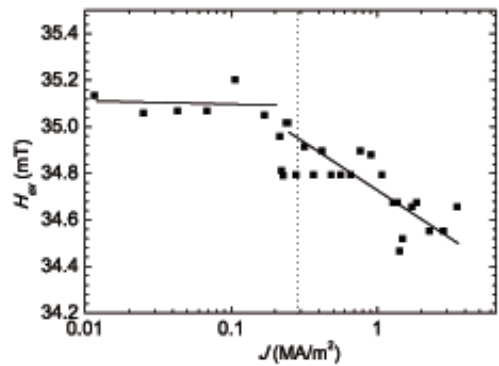}\\
\caption{The current density $J$ dependence of exchange bias $(H_{ex})$
that extracted from the $R-\mu_{0}H$ curves at different voltages (0.1 - 2.2 V)
for MTJ1. The position of the dot line corresponds to the beginning of the FN tunneling ($V$=1.35V). The solid lines are guide to the eye.} \label{fig3}
\end{figure}

We use the following equations to evaluate the thermal fluctuation
of the barrier height due to the current induced heat and the corresponding
thermal temperature across the barrier. For an MTJ, the temperature
dependence of the junction resistance is usually explained in terms
of elastic and inelastic tunneling\cite{34CHShang}. The temperature dependence of the averaged conductance in the PC
and APC states of MTJs is given by the following equations without
considering the inelastic tunneling here \cite{34CHShang}:

\begin{equation}\label{eq1}
\begin{split}
     G_{PC} = & G_{T}[1 + P_{1}P_{2}]  \\
    G_{APC} = & G_{T}[1 - P_{1}P_{2}] ,
\end{split}
\end{equation}

where $G_{T}=G_{0}CT/sin(CT)$, $G_{0}$ is the conductance of the MTJ
at zero temperature; $C=1.39\times10^{-4}t/(\phi^{1/2})$, with the
barrier thickness $(t)$ in angstroms and the barrier height $(\phi)$
in electron-volts; $P_{1}$ and $P_{2}$ are effective spin
polarizations of two ferromagnetic electrodes. Here we use $\delta R/R$ to define the shift
magnitude, and the relation $\delta \phi$ = n $k_{B}\delta T$ to obtain thermal temperature $\delta T$ across the
barrier. It is found that the relation between $\delta \phi/\phi$ and
$\delta R/R$ in the PC and APC states follows by

\begin{equation}\label{eq2}
 \delta \phi/\phi \approx 2(\delta R/R),
\end{equation}

assuming $sin(CT) \approx sin((C+\delta C)(T+\delta T))$ and $(T+\delta T)/T \approx
1$ because $\delta C$ and $\delta T$ are small. $\delta \phi$ is the barrier height change due to the thermal
fluctuation. For calculation, $\phi_{PC}$ ($\phi$ in the PC state) and $\phi_{APC}$ ($\phi$ in the APC state) at different voltages are fitted by the Brinkman's model\cite{30WFBrinkman}, which mainly deals with
trapezoidal barriers in MTJs.

According to Eq. (1), we obtain $\delta V$ = $V(\delta R/R)/2$ in the PC and APC states. Here the spin-dependent Seebeck coefficient (S) is defined as $\delta V/\delta T$. Using $\delta T$ and $\delta V$ mentioned above, we obtain

\begin{equation}\label{eq3}
 S = \delta V/\delta T \\
   = (n/4)k_{B}(V/\phi).
\end{equation}

It is obvious that S is independent of the $\delta R/R$ ratio.

Fig. 4 (a) shows the current density dependence of the thermal
temperature $\delta T$ in the PC state for MTJ1 before and after the FN
tunneling occurs. There are three distinct regions of $\delta T$
changing with the current density. Besides two cases (Case1 and Case2) given above, $\delta T$ in the normal tunneling regime (Case3) close to Case1 is also plotted for comparison. With the increase of current in Case3, the heat increases gradually. When the FN tunneling occurs in Case1, the heat decreases due to the increase of transparency of the barrier. However, once the spacer-like barrier becomes obvious in
Case2, $\delta T$ increases again because of the strong scattering. A $\delta T$ range of 0.2 - 10 K is observed in Case1 and Case2, while it changes from 3.5 K to 10 K in the normal tunneling regime. Several tens or hundreds mK of $\delta T$ have been observed in MTJs \cite{8NLiebing, 10WWLin}. The thermovoltage $\delta V$ is plotted as a function of the current density in Fig. 4 (b). The corresponding $\delta V$ value ranges from 0.15 to 1.6 mV in Case1 and Case3, while it reaches 17.1mV in Case2. Clearly, $\delta V$ varies similarly with the
current density compared with $\delta T$ due to the same power-law of $\delta R/R$. Both $\delta T$ and $\delta V$ reaches a maximum after the FN tunneling occurs, which may indicate the increase of transparency of the barrier is not obvious at the beginning of the FN tunneling. Besides the spin-dependent contribution, the resistance shift ($\delta R$) includes the non spin - dependent part, both together may be responsible for $\delta T$ and $\delta V$ shown in Fig.4.


\begin{figure}
\includegraphics[width=8cm]{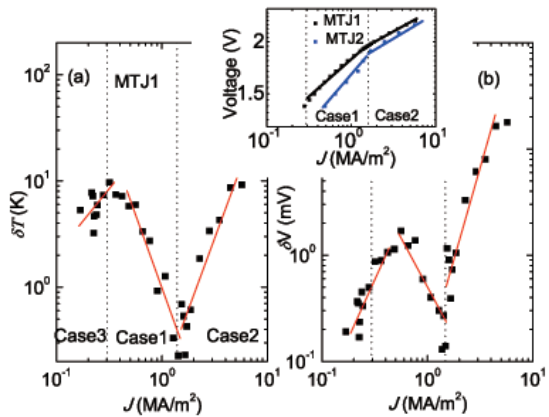}\\
\caption{The current density dependence of $\delta T$ (a) and
$\delta V$ (b) in the PC state for MTJ1 ($V$ = 1.0 - 2.2 V). The
inset plots the \emph{V }- \emph{J} curve after the FN
tunneling occurs for MTJ1 and MTJ2. The dot lines in Fig.4 and in the insert show the regions of three Cases (Cases1-3).} \label{fig4}
\end{figure}

The spin-dependent thermal parameters normally scale linearly with
current ($\propto J$) in MTJs \cite{10WWLin}, spin valves \cite{11HMYu} or ferromagnet/metal
contact\cite{12ASlachter}. Both $\delta T$ and $\delta V$ show a
roughly linear response in different regimes (Cases 1-3) for MTJ1. This linear relation permit us to estimate the spin-dependent Seebeck coefficient $S$ in these MTJs. The $S$ values in the PC and APC states for MTJ1 with power are given in Fig. 5 (a) (square data points). For the S calculation, we use Eq.(3) by selecting n = 1, and more discussion about n will be given below. For comparison, S in the PC state for MTJ2 is also shown (diamond data points). It is found that both MTJ1 and MTJ2 behave similarly. Because the barrier height cannot be precisely evaluated after the tunneling barrier becomes spacer-like, here we focus on the results in Case1 and Case3. It is found S and dS increase linearly with power in two cases. Here dS is the difference in S between APC
and PC states, as shown in Fig.5 (b). Before more electrons appear in the tunneling barrier, S is relatively small in two resistance
states, several tens of $\mu V/K$, which is similar to that in many cases \cite{5MWalter,
6ASlachter, 7MCzerner, 8NLiebing, 10WWLin,14ZHZhang}. At the beginning of the FN tunneling, the lowest S values of 55 $\mu V/K$ in the PC state and of 83 $\mu V/K$ in the APC state are obtained. Once the FN tunneling is dominant, S increases fast with power until the spacer-like barrier forms. The maximum values of 1.72 mV/K in the APC state and 0.91 mV/K in the PC
state are obtained for MTJ1. Obviously, the magnitude of S can be tuned in these MTJs by the applied power, which is due
to the variation in the transparency of the barrier in our case. A value of dS ranging from 9 to 1190 $\mu V/K$ is obtained for MTJ1 (see center square data points in Fig. 5 (b)). The S and dS values obtained in the previous reports \cite{7MCzerner,8NLiebing, 10WWLin,14ZHZhang} are within the range as shown in Fig.5.

Here we discuss the n value as shown in Eq. (3). The Fano effect appears due to the quantum interferences of the waves resonantly transmitted through
a discrete level and those transmitted nonresonantly through continuum of states\cite{35PTrocha}. Such effect is usually accompanied by shot noise\cite{36FNHooge, 37RJSchoelkopf}, which is due to the current through the device fluctuating around its average value. The shot noise is frequency independent, and its total energy $E$  is (2Fe$I$)$R$ (=2Fe$V$) \cite{36FNHooge, 37RJSchoelkopf}, where e is the electron charge and F is the Fano facor. For simplicity, we select F = 1 here. In our case, the energy $\delta E$ related to $\delta T$ and $\delta V$ can be given by (2eI)$\delta R$, which corresponds to the thermal energy part in the current-induced fluctuation. If using $I = V/R$, then $\delta E$ is given by

\begin{equation}\label{eq4}
 \delta E = 2eV (\delta R/R).
\end{equation}

Like $\delta \phi$, if we use the relation $\delta E = n k_{B}\delta T$ to define $\delta T$, then we obtain S as follows

\begin{equation}\label{eq5}
 S =  (n/4)(k_{B}/e).
\end{equation}

Here S is proportional to the unit of $k_{B}/e$, which is independent of voltage and barrier height. This is a character of quantum interference\cite{25OKarlstrom, 35PTrocha}. Because of n $>$ 1 when the FN tunneling occurs (see the discussion below), $k_{B} \delta T$ $<$ $\delta E$ (n $k_{B}\delta T)$ $<<$ $E$ (2e$V$), which satisfies with quantum limit \cite{24CWJBeenakker}. The barrier with asymmetric barrier heights in our MTJs can be as a qusi-quantum well especially after the FN tunneling appears, and the oscillation of TMR shown in Fig.2 (a) further proves it. In this case, the discrete levels (barrier heights) change with the applied voltage.

Without the quantum interference effect, S is small as shown in Case3 (Fig. 5). Different S obtained by Eq.(5) may imply an ability to carry electrons, which is related to $ eV$. Actually if we assume $ eV= \phi$, Eq.(5) is back to Eq.(3), which suggests Eq.(3) is correct only around the beginning of the FN tunneling regime where the barrier height is equal to $eV$. Moreover, we may use Eq.(5) to evaluate the n value in these MTJs since the variation of barrier height can be equal to that of voltage. It is obtained n increases from around 1 to 80, as the two black dash lines shown in Fig. 5 (a). Because of the Fano factor F $>$ 1 in the quantum regime\cite{38GIannaccone, 39AThielmann}, the maximum of (n/4) in our case will be less than 20. A value of (n/4) up to 10 has been given in Ref.\cite{25OKarlstrom}.

According to Eq.(5), S is charge dependent. If considering the contribution of spin-up($\uparrow$) and spin-down ($\downarrow$) electrons, S is equal to  $(S_{\uparrow}+S_{\downarrow})/2$ in the two resistance states. Because the TMR is lower than $1.5\%$ in Case1 and the corresponding spin polarization is less than $8.6 \%$, the contributions from majority and minority electrons are similar for the Seebeck coefficient S in these MTJs. $S_{\uparrow}$ and $S_{\downarrow}$ are almost half of the S value. Compared to the PC state, more scattering occurs in the APC state, which may be responsible for the big dS in Case1 in these MTJs.


\begin{figure}
\includegraphics[width=8cm]{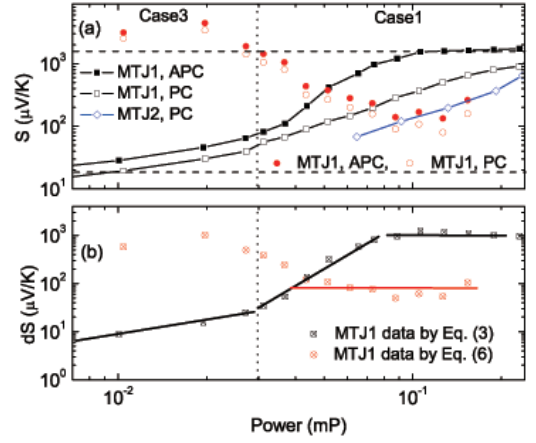}\\
\caption{(a) The power dependence of spin-dependent Seebeck coefficient in Case1 and Case3 at two resistance states for MTJ1 and MTJ2, (b) the dS values as a function of power for MTJ1. The corresponding voltage in (a) and (b) is 1.0 - 2.2 V. The square and diamond data points are obtained using Eq. (3). The circle data points are given by Eq. (6). The range between two black dash lines in (a) shows the variation of S decided by Eq. (5). The solid lines in (b) are guide to the eye.} \label{fig5}
\end{figure}

Finally we use another method shown in Ref.\cite{14ZHZhang} to calculate the values of S and dS for MTJ1. As suggested by Ref.\cite{14ZHZhang}, to see the Seebeck effect, a linear relation in the voltage versus current density curve should be satisfied (the inset of Fig.4). The thermovoltage $dV$ in this case follows:

\begin{equation}\label{eq6}
dV = S \sum(\eta R_{\kappa}R)I^{2},
\end{equation}

here the thermal parameter $\eta$ is asymmetric, and the heat resistance $R_{\kappa}$ equals to
$d/(\kappa A)$ with thermal conductivity $\kappa$, cross-sectional
area $A$ and the total thickness $d$ of the MTJ device. Because the
resistance mainly comes from the insulating barrier in these MTJs, we set $R =
R_{PC}$ in the PC state and $R_{APC}$ in the APC state. $\eta$ in our case is bias -
dependent, but we use a constant $\eta$ (0.5) (see Fig.1). $\kappa$ = 1.5 W/(Km) is used for $AlO_{x}$
during the calculation \cite{40YSJu}. The calculated results for MTJ1 in the PC and APC states
are given in Fig.5 (red open and solid circle data points). Some deviation of S in two resistance states appears in the low power range because the voltage - current curve loses its linear relation (see the insert of Fig.4). The values of S and dS obtained by Eq. (6) are in the range of those obtained by Eq. (3) and Eq. (5). However, S and dS obtained in this way do not alter much with power, as indicated by the red solid line shown in Fig. 5 (b).

\section{Conclusion}

In this work, we take the advantage of MTJs with asymmetric oxidized Al-oxide. In the FN tunneling regime,
MTJs are turned to be an oscillatory TMR effect which is reversible by voltage.
The thermal fluctuation on exchange bias is observed together with FN tunneling process. Such fluctuation is
induced by current induced heat around FN tunneling
regime, and the fluctuation temperature increases with the applied voltage.

Furthermore, the thermal transports have been investigated in these MTJs. The tunneling barrier itself can serve as a stable thermal source in MTJs, which supplies a direct heat source for spin-dependent thermoelectric studies. It is the current-induced heat caused by quantum interference that makes the thermoelectric phenomena. The linear relation between the thermal temperature, thermovoltage and the applied current proves the thermoelectric nature. The Seebeck coefficient in these MTJs can be tuned and large, which suggests that MTJs may be a good candidate for thermoelectric application.

\section{Acknowledgment}
This project was supported by the State Key Project of Fundamental Research of Ministry of Science and
Technology [MOST, Project  No. 2010CB934400 and No. 2011YQ120053] and National Natural Science Foundation [NSFC, Grant No. 11104338 and 51021061].

\end{document}